\documentclass[smallextended]{svjour3}

\journalname{ARMA}

\usepackage{mathptmx}    

\usepackage{amsmath}

\usepackage{amssymb}

\usepackage{mathtools}                                 

\newcommand{\bfI}{\mathbb{I}}
\newcommand{\bfQ}{\tens{Q}}

\newcommand{\bfzero}{\tens{0}}

\newcommand{\calF}{\mathcal{F}}

\newcommand{\ehat}{\vec{e}}
\newcommand{\lhat}{\vec{l}}
\newcommand{\nhat}{\vec{n}}

\newcommand{\bfu}{\tens{u}}                  

\newcommand{\eps}{\varepsilon}

\newcommand{\Sminus}{S_{-}}
\newcommand{\Splus}{S_{+}}
\newcommand{\Spm}{S_{\pm}}

\newcommand{\SNI}{S_{\text{NI}}}

\newcommand{\Stildepm}{\widetilde{S}_{\pm}}

\newcommand{\overbar}[1]{%
  \mkern1.5mu\overline{\mkern-1.5mu#1\mkern-1.5mu}\mkern1.5mu}

\newcommand{\Vbar}{%
  \mkern1.0mu\overline{\mkern-1.0mu{V}\mkern-1.0mu}\mkern1.0mu}
\newcommand{\xbar}{%
  \mkern1.5mu\overline{\mkern-1.5mu{x}}}
\newcommand{\gradbar}{%
  \mkern1.0mu\overline{\mkern-1.0mu\nabla\mkern-1.0mu}\mkern1.0mu}

\newcommand{\Qbar}{%
  \mkern1.5mu\overline{\mkern-1.5mu\bfQ\mkern-1.5mu}\mkern1.5mu}
\newcommand{\xibar}{%
  \mkern1.5mu\overline{\mkern-1.5mu\xi\mkern-1.5mu}\mkern1.5mu}

\newcommand{\xibbar}{\xibar_{\text{b}}}
\newcommand{\xiNIbar}{\xibar_{\text{NI}}}

\newcommand{\Omegabar}{%
  \mkern2.5mu\overline{\mkern-2.5mu\Omega\mkern-2.0mu}\mkern2.0mu}
\newcommand{\tOmegabar}{%
  \mkern3.0mu\overline{\mkern-3.0mu\Omega\mkern-2.5mu}\mkern2.5mu}

\newcommand{\Fbar}{%
  \mkern4.5mu\overline{\mkern-4.5mu\calF\mkern-1.0mu}\mkern1.0mu}

\newcommand{\sQbar}{\mkern1.5mu\overline{\mkern-1.5mu{Q}}}

\newcommand{\jbar}{%
  \mkern1.5mu\overline{\mkern-1.5mu{j}\mkern1.0mu}\mkern-1.0mu}
\newcommand{\kbar}{\overline{k}}

\newcommand{\Lbar}{\overline{L}}

\newcommand{\Fxi}{F_{\xi}}
\newcommand{\Fbarxibar}{\overbar{F}_{\!\xibar}}

\newcommand{\Ftilde}{%
  \mkern4.5mu\widetilde{\mkern-4.5mu\calF\mkern2.0mu}\mkern-2.0mu}

\newcommand{\Qtilde}{\widetilde{\bfQ}}
\newcommand{\Qtildeminus}{\widetilde{\bfQ}_{-}}
\newcommand{\Qtildeplus}{\widetilde{\bfQ}_{+}}
\newcommand{\Qtildepm}{\widetilde{\bfQ}_{\pm}}

\newcommand{\fb}{f_{\text{b}}}
\newcommand{\fe}{f_{\text{e}}}

\newcommand{\fbepstilde}{\tilde{f}_{\text{b},\eps}}
\newcommand{\fbztilde}{\tilde{f}_{\text{b},0}}

\newcommand{\ASC}{A_{\text{SC}}}
\newcommand{\ANI}{A_{\text{NI}}}
\newcommand{\ASH}{A_{\text{SH}}}

\newcommand{\TSC}{T_{\text{SC}}}
\newcommand{\TNI}{T_{\text{NI}}}
\newcommand{\TSH}{T_{\text{SH}}}

\newcommand{\xib}{\xi_{\text{b}}}
\newcommand{\xin}{\xi_{\text{n}}}
\newcommand{\xiNI}{\xi_{\text{NI}}}

\DeclareMathOperator{\diam}{diam}
\DeclareMathOperator{\tr}{tr}

\newcommand{\myoverbracket}[1]{\overbracket[.1ex][.5ex]{#1}}

\begin{document}

\title{Scalings and Limits of Landau-de\,Gennes Models \\
  for Liquid Crystals\thanks{Supported in part by U.S.\ National
    Science Foundation grant DMS-1211597}}

\subtitle{A Comment on Some Recent Analytical Papers}


\author{Eugene C. Gartland, Jr.}

\institute{E. C. Gartland, Jr. \at
  Department of Mathematical Sciences,
  Kent State University,
  Kent, Ohio 44242, USA \\
  Tel.: +1-330-672-9112,
  Fax: +1-330-672-2209,
  \email{gartland@math.kent.edu}}



\date{\today}

\maketitle

\begin{abstract}
  Some recent analytical papers have explored limiting behaviors of
  Landau-de\,Gennes models for liquid crystals in certain extreme
  ranges of the model parameters: limits of ``vanishing elasticity''
  (in the language of some of these papers) and ``low-temperature
  limits.''  We use simple scaling analysis to show that these limits
  are properly interpreted as limits in which geometric length scales
  (such as the size of the domain containing the liquid crystal
  material) become large compared to intrinsic length scales (such as
  correlation lengths or coherence lengths, which determine defect
  core sizes).  This represents the natural passage from a mesoscopic
  model to a macroscopic model and is analogous to a ``London limit''
  in the Ginzburg-Landau theory of superconductivity or a ``large-body
  limit'' in the Landau-Lifshitz theory of ferromagnetism.  Known
  relevant length scales in these parameter regimes (nematic
  correlation length, biaxial coherence length) can be seen to emerge
  via balances in equilibrium Euler-Lagrange equations associated with
  well-scaled Landau-de\,Gennes free-energy functionals.

\end{abstract}

\section{Introduction}

The Landau-de\,Gennes and Oseen-Frank models are the two most widely
used continuum models to characterize equilibrium orientational
properties of materials in the nematic liquid crystal phase.  The
Landau-de\,Gennes model is a mesoscopic phenomenological model
expressed in terms of an integral functional of a tensor field $\bfQ$,
the ``tensor order parameter.''  In its simplest form (the ``equal
elastic constant model''), the functional can be written
\begin{subequations}\label{eqn:FQ}
\begin{equation}\label{eqn:FQa}
  \calF[\bfQ] = \int_{\Omega}
  \Bigl[ \frac{L}2 | \nabla\bfQ |^2 + \fb(\bfQ) \Bigr] \, \text{d}V .
\end{equation}
Here $\calF$ gives the free energy of a material occupying the region
$\Omega$, and $\fb$ represents the free-energy density of a
homogeneous bulk material, which in general is given in the form
\begin{equation}
  \fb(\bfQ) = \frac{A}2 \tr(\bfQ^2) - \frac{B}3 \tr(\bfQ^3) +
  \frac{C}4 \tr(\bfQ^2)^2 .
\end{equation}
\end{subequations}
The parameters $L$, $A$, $B$, and $C$ are material dependent, with
$L$, $B$, and $C$ positive and $A$ (which can be positive, negative,
or zero) usually taken to have a simple linear dependence on
temperature:
\begin{equation}\label{eqn:AofT}
  A = a_0 ( T - \TSC ) , \quad a_0 > 0 .
\end{equation}
Here $T$ denotes temperature, and $\TSC$ represents the
``super-cooling temperature'' (sometimes denoted $T^*$\!\!, discussed
below).  Stationary points of $\calF$ give equilibrium orientational
states, with the global minimum determining the structural phase of
the system for a given set of boundary conditions and parameter
values.

The $\bfQ$ tensor is formally defined as the traceless part of the
second-moment tensor of the orientational probability distribution
function (which, if known, would contain complete information about
the orientational state at a point):
\begin{equation*}
  \bfQ = \bigl\langle \lhat\otimes\lhat \bigr\rangle - \frac13 \bfI .
\end{equation*}
Here the unit vector $\lhat$ denotes the direction of the
distinguished axis of the anisometric molecular architecture (usually
the long axis of elongated, rod-like molecules), and $\bfI$ is the
identity tensor.  As such, $\bfQ$ is real, symmetric, traceless, and
generically would have three distinct real eigenvalues and associated
orthogonal eigenvectors, representing a ``biaxial'' state of order.
The eigenvectors of $\bfQ$ provide information about the directions of
orientational ordering at a point, while the eigenvalues give
information about the degrees of order---note that the value of $\fb$
depends only on the eigenvalues of $\bfQ$.  An isotropic (totally
disordered) state corresponds to $\bfQ = \bfzero$ (the zero tensor).
For given $B,C>0$, the shape of the surface of $\fb$ as a function of
the two independent eigenvalues of $\bfQ$ changes as temperature
changes.  For $T > \TSH > \TSC$, where $\TSH$ denotes the
``super-heating temperature,'' $\bfQ=\bfzero$ is the global minimizer
of $\fb$, and no ordered phase exists---$\fb$ has no other stationary
points.  For $T < \TSC$, the isotropic phase ceases to be locally
stable, and the only minimizer of $\fb$ is a $\bfQ$ tensor with a
dominant eigenvalue and a degenerate pair of equal eigenvalues.  Since
$\tr(\bfQ)=0$, such a $\bfQ$ can be written
\begin{equation}\label{eqn:uniaxialQ}
  \bfQ = S \Bigl( \nhat\otimes\nhat - \frac13 \bfI \Bigr) .
\end{equation}
Here $S$ is referred to as the ``scalar order parameter,'' and $\nhat$
is the distinguished eigenvector.  This is known as a ``uniaxial''
state of order.  For $T$ in the narrow range between $\TSC$ and
$\TSH$, both the isotropic and the uniaxial phases are locally stable,
and there is a first-order phase transition at the ``nematic-isotropic
transition temperature'' $\TNI$, below which the ordered phase becomes
the global minimizer.  Simple relationships among $A$, $B$, and $C$
are associated with these temperatures:
\begin{equation}\label{eqn:temps}
  \TSC < \TNI < \TSH ~~ \leftrightarrow ~~
  \ASC = 0 < \ANI = \frac{B^2}{27C} < \ASH = \frac{B^2}{24C} .
\end{equation}
Every material that admits a nematic liquid crystal phase would reside
in such a phase for some finite range of temperatures below $\TNI$;
Landau-de\,Gennes models (which are formally derived via truncated
expansions) contain no parameters that indicate the extent of this
range.

The distortional elastic term associated with the constant $L$ in
\eqref{eqn:FQa} penalizes spatial variations of the $\bfQ$ tensor,
while the bulk ordering potential $\fb$ strives to put the eigenvalues
of $\bfQ$ into certain wells.  More complete versions of this model
can contain additional terms incorporating more elastic constants,
more bulk constants, terms associated with couplings to electric or
magnetic fields, terms associated with chirality, flexoelectric
effects, ferroelectric effects, surface anchoring potentials, and
more.  See \cite{mottram:newton:14} for a gentle introduction to the
model or \cite{degennes:prost:93,sonnet:virga:12} for textbook
treatments.  The form \eqref{eqn:FQ} is the simplest form of the model
and is sufficient for the purposes at hand.  Landau-de\,Gennes models
occupy a similar position in the theory of liquid crystals to that of
Ginzburg-Landau models for the theory of superconductivity and
Landau-Lifshitz models for ferromagnetism.

The Oseen-Frank model is a macroscopic phenomenological model for
liquid crystal orientational properties.  It is expressed in terms of
an integral functional of a unit-length vector field $\nhat$ (the
``director field'').  In its simplest (``equal elastic constant'')
form, the functional can be written
\begin{equation}\label{eqn:OZF}
  F[\nhat] = \frac{K}2 \int_{\Omega} | \nabla\nhat |^2 \, \text{d}V .
\end{equation}
Here $F$ gives the distortional elastic energy of the material
occupying $\Omega$, and $\nhat$ represents the average orientation of
the distinguished axes of the molecules in a fluid element at a point
(and can be identified with the eigenvector associated with the
distinguished eigenvalue of a uniaxial $\bfQ$ tensor, as in
\eqref{eqn:uniaxialQ}).  The elastic constant $K$ is a
material-dependent and temperature-dependent parameter.  Equilibrium
orientational states are given by stationary points of $F$ constrained
by boundary conditions and the pointwise unit-length constraint on
$\nhat$, with the structural phase of the system (for a given $K$ and
boundary conditions) again given by the global minimizer.  As is the
case with the Landau-de\,Gennes free energy $\calF$, more realistic
models for $F$ involve many more terms, parameters, couplings, and
effects.  Standard references include
\cite{chandrasekhar:92,degennes:prost:93,stewart:04,virga:94}.
Oseen-Frank models have been widely and successfully used for many
years to model liquid crystal systems at the scales of typical devices
and experiments.

These two models differ in several ways.  The Landau-de\,Gennes model
allows for both spatially varying degrees of order and biaxiality, and
it is expressed in terms of a tensor field.  The Oseen-Frank model, on
the other hand, assumes a uniform degree of orientational order, as
well as a uniaxial state of order, and is expressed in terms of a
vector field.  Another difference between the models is that common
liquid crystal ``defects'' (such as ``point defects'' and
``disclination lines'') are singularities of the director field in the
Oseen-Frank model, whereas in the Landau-de\,Gennes model, these have
a small but finite ``core size'' and the associated tensor field
$\bfQ$ remains smooth throughout a neighborhood of such a defect.  In
Oseen-Frank free-energy functionals, such as \eqref{eqn:OZF}, point
defects have finite free energy, while the free energy of a
disclination line is infinite.  In the Landau-de\,Gennes model,
however, the free energies of both are finite.  In general, defects
are caused by the conflicting demands of distortional elasticity,
boundary conditions, external electric or magnetic fields, and the
like.  Landau-de\,Gennes models are typically employed in problems in
which geometric length scales are not too large compared to intrinsic
length scales (such as defect core sizes), while Oseen-Frank models
are used when the geometric length scale is much larger than the core
size (and the fine-detail structure of the core is not important).

In recent years, Landau-de\,Gennes models have received considerable
attention from the mathematical analysis community.  In particular,
three papers have appeared in this journal that rigorously explore
(among other issues) limiting behaviors of this model for certain
extreme ranges of the elastic constants
\cite{bauman:park:phillips:12,golovaty:montero:14,majumdar:zarnescu:10}.
Analyses of somewhat related limits (emphasizing behavior at low
temperatures) are found in
\cite{contreras:lamy:14,majumdar:pisante:henao:15}.  Our main purpose
here is to explain how these limits should be interpreted.

In \cite{majumdar:zarnescu:10} (the earliest of these papers), the
model is taken in the form \eqref{eqn:FQ} above, and the authors
motivate their work as follows: ``we study the limit of vanishing
elastic constant $L\rightarrow0$ \ldots\ the limit $L\rightarrow0$ is
a physically relevant limit since the elastic constant $L$ is
typically very small, of the order $10^{-11}$J/m.''  In
\cite{bauman:park:phillips:12}, a slightly more general form of the
model is used:
\begin{subequations}\label{eqn:Feps}
\begin{equation}\label{eqn:Fepsa}
  \calF_{\eps}[\bfQ] = \int_{\Omega}
  \bigl[ \fe( \nabla\bfQ ) + \eps^{-2} \fb( \bfQ ) \bigr] \, \text{d}V ,
\end{equation}
where
\begin{equation}
  \fe(\nabla\bfQ) =
  \frac{L_1}2 Q_{ij,k} Q_{ij,k} +
  \frac{L_2}2 Q_{ij,j} Q_{ik,k} +
  \frac{L_3}2 Q_{ij,k} Q_{ik,j} .
\end{equation}
\end{subequations}
Here $Q_{ij}$ are the components of $\bfQ$ with respect to a fixed
Cartesian frame; $Q_{ij,k}$ denotes $\partial Q_{ij} / \partial x_k$;
and summation over repeated indices is implied.  The dimensionless
parameter $\eps$ is artificially introduced in order to be able to
drive the elastic constants $L_1$, $L_2$, and $L_3$ to zero
simultaneously.  The $L_1$ term above corresponds to the $L$ term in
\eqref{eqn:FQa}.  The authors motivate their work ``Our goal in this
paper is to investigate minimizers [of \eqref{eqn:Feps}] and to
analyze their behavior in the vanishing elastic energy limit, $\eps
\rightarrow 0$.''

In these two papers, the models are analyzed in fully dimensional
form.  In \cite{golovaty:montero:14}, a dimensionless model is
studied, and it is expressed in terms of the second-moment tensor
$\bfu = \langle \lhat\otimes\lhat \rangle$ (instead of $\bfQ$).  The
model there is written
\begin{subequations}
\begin{equation}\label{eqn:Eepsa}
  E_{\eps}[\bfu] = \int_{\Omega}
  \Bigl[ \frac12 |\nabla\bfu|^2 + \frac1{~\eps^2} W(\bfu) \Bigr]
  \, \text{d}V ,
\end{equation}
where
\begin{equation}
  W(\bfu) = \frac12 \tr \bigl( ( \bfu - \bfu^2 )^2 \bigr) .
\end{equation}
\end{subequations}
The bulk ordering potential $W$ here is constructed by design to have
a minimizer (in the class $\tr(\bfu)=1$) at a perfectly ordered
uniaxial state $\bfu = \nhat\otimes\nhat$, with $|\nhat|=1$ but with
the direction of $\nhat$ arbitrary, and $\eps$ is referred to as a
``dimensionless elastic constant.''  With the help of the relation
$\bfu = \bfQ + \frac13 \bfI$, the model can be identified with a
certain constrained form of \eqref{eqn:FQ}.  Again, the limit $\eps
\rightarrow 0$ is explored.

All three of these papers explore similar limits, related to
``vanishing elasticity,'' and are influenced by similar analyses of
Ginzburg-Landau models found in
\cite{bethuel:brezis:helein:93,bethuel:brezis:helein:94} and
elsewhere.  All three obtain (away from a singular set) limiting
uniaxial minimizers of the form \eqref{eqn:uniaxialQ}, with constant
$S$ determined so as to provide a minimum of $\fb$, and with the
director field $\nhat$ corresponding to the minimizer of an
appropriate Oseen-Frank model.  We show that these limits are properly
interpreted not as limits of vanishing elasticity but as limits in
which intrinsic length scales (associated with defect core sizes and
such) become vanishingly small compared to geometric length scales
(associated with the size of the problem domain $\Omega$).

The limits analyzed in
\cite{contreras:lamy:14,majumdar:pisante:henao:15} pertain to the
behavior of Landau-de\,Gennes models at low temperatures, where the
changing landscape of the bulk free-energy $\fb$ penalizes biaxial
order less and at the same time shrinks defect core sizes.  This ``low
temperature'' limit as well corresponds to a ``zero core size'' limit
(with additional features), and the close relationship between these
two limits is demonstrated below.  Different length scales are known
to be associated with defect core sizes in these two different
regimes, and it is also shown below how these can be identified via
balances in appropriate scalings of the Euler-Lagrange equations
associated with \eqref{eqn:FQ}.

\section{Scaling analysis}

\label{sec:analysis}

Values of the material parameters in \eqref{eqn:FQ} are usually found
quoted in SI units.  For a somewhat typical material, they are roughly
in the following ranges:
\begin{equation}\label{eqn:LABCapprox}
  L \approx 10^{-11}\text{J/m} , \quad
  A, B, C \approx 10^5\,\text{J/m}^3 .
\end{equation}
See for example \cite[Table\,1, p.\,168]{priestly:wojtowicz:sheng:75}.
We observe that the numerical value of $L$ is 16 orders of magnitude
smaller than the values of $A$, $B$, and $C$ when expressed in these
units.  If, however, lengths are expressed in units of nanometers
(instead of meters), these values become
\begin{equation*}
  L \approx 10^{-20}\text{J/nm} , \quad
  A, B, C \approx 10^{-22}\text{J/nm}^3 .
\end{equation*}
Now the value of $L$ is two orders of magnitude \emph{larger} than
$A$, $B$, and $C$.  The point is that the elastic constants and the
bulk constants have different physical dimensions (energy per unit
length versus energy per unit volume) and can't be compared.  To
determine what is ``big'' versus what is ``small,'' one must
non-dimensionalize.  An additional point to be made is that as $L
\rightarrow 0$, with the model left in dimensional form (as in
\cite{bauman:park:phillips:12,majumdar:zarnescu:10}), one quickly gets
beyond the measured values of $L$ for real liquid crystal materials.

\subsection{Large-body limit}

There are numerous ways to non-dimensionalize \eqref{eqn:FQ}, and
these depend in general on the particular problem at hand.  In
essence, in order to form an appropriate dimensionless coupling
coefficient between the elastic terms and the bulk terms in the
free-energy density, one requires a characteristic elastic constant, a
characteristic bulk constant, and a characteristic length scale from
the geometry of the problem domain.  For the purpose of understanding
the ``vanishing elasticity'' limit, a simple rescaling can be done as
follows.  Let $R$ denote a characteristic geometric length scale.  For
example, this could be the radius or diameter of a liquid crystal
droplet or capillary, or the cell gap of a liquid crystal thin film.
Rescale lengths by $R$,
\begin{subequations}\label{eqn:xbar}
\begin{equation}
  \xbar_i = \frac{x_i}{R} , \quad
  R = \diam(\Omega),
\end{equation}
so that
\begin{equation}
  \nabla = \frac1R \gradbar , \quad
  \text{d}V = R^3 \text{d}\Vbar , \quad
  \diam(\Omegabar) = 1 .
\end{equation}
\end{subequations}
Let $\ANI$ denote the value of the $A$ parameter at $T = \TNI$, as in
\eqref{eqn:temps}, and take this as our characteristic bulk
parameter---other natural candidates for this would be $\ASH$ or
$B^2\!/C$, or simply $B$ or $C$.  The tensor order parameter $\bfQ$ is
dimensionless by definition, however it is convenient to rescale it as
well, as this allows one to eliminate another parameter from the
model.  Thus we take
\begin{equation*}
  \bfQ = \alpha \, \Qbar , \quad
  \alpha := \frac1{\sqrt{27}} \frac{B}{C} .
\end{equation*}
Other multiples of $B/C$ would work equally well.  After simplifying,
we obtain
\begin{subequations}\label{eqn:FQbar}
\begin{equation}\label{eqn:FQbara}
  \Fbar[\Qbar] = \int_{\tOmegabar} \Bigl[
  \frac12 \xiNIbar^2 \bigl|\gradbar\Qbar\bigr|{}^2 +
  \frac{\theta}2 \tr\bigl(\Qbar^2\bigr) -
  \sqrt{3} \tr\bigl(\Qbar^3\bigr) +
  \frac14 \tr\bigl(\Qbar^2\bigr){}^{\!2} \Bigr] \text{d}\Vbar ,
\end{equation}
where
\begin{equation}\label{eqn:FQbarb}
  \Fbar = \frac{\calF}{\alpha^2 \ANI R^3} , \quad
  \xiNIbar = \frac{~\xiNI}{R} , \quad
  \xiNI := \sqrt{ \! \frac{L}{\,\ANI} } , \quad
  \theta := \frac{A}{~\ANI} = \frac{T-\TSC}{\TNI-\TSC} .
\end{equation}
\end{subequations}

The parameter $\xiNI$ can be interpreted as the ``nematic correlation
length'' at the nematic-isotropic transition temperature $T=\TNI$ and
$\theta$ as a reduced temperature, with the corresponding values
\begin{equation*}
  T = \TSC, \TNI, \TSH ~~ \leftrightarrow ~~ \theta = 0, \, 1, \, 9/8 .
\end{equation*}
Here and in \eqref{eqn:FQbar}, we have used the relations for $\ANI$
and $\ASH$ in \eqref{eqn:temps}.  In a typical liquid-crystal system
(for example, a thin film of cell gap $R$ containing a
low-molecular-weight liquid crystal material), we could have
\begin{equation*}
  \xiNI \approx 10\,\text{nm} , ~ R \approx 10\,\mu\text{m}
  ~~ \Rightarrow ~~ \xiNIbar \approx 10^{-3} .
\end{equation*}
The approximate value for $\xiNI$ above can be seen to follow roughly
from \eqref{eqn:LABCapprox}:
\begin{equation*}
  \xiNI^2 = \frac{L}{~\ANI} \approx
  \frac{10^{-11}\text{J/m}}{10^5\,\text{J/m}^3} =
  10^{-16}\text{m}^2 .
\end{equation*}
Such small values of the dimensionless coupling coefficient
($\xiNIbar^2 \approx 10^{-6}$) would place a large weight on the terms
from the bulk ordering potential, strongly encouraging uniaxial order
at the bulk-minimizing value of the scalar order parameter.

The reduced temperature $\theta$ corresponds to the temperature above
or below $\TSC$ measured in units of $(\TNI-\TSC)$, which is
approximately the width of the coexistence temperature region (where
both the isotropic and the ordered phases are locally stable).  For a
fairly pure sample of a low-molecular-weight liquid crystal, the
coexistence region would typically be one or two degrees Celsius wide,
and so one can think of $\theta$ as roughly the temperature
above/below $\TSC$ in degrees Celsius.  The critical temperatures of
liquid crystal materials vary quite a lot.  The values of $\TNI$ for
three commonly studied liquid crystals are given in \cite[Table~D.3,
p.\,330]{stewart:04} as $\TNI = 35.4^{\circ}$C (5CB), $45.1^{\circ}$C
(MBBA), $135.5^{\circ}$C (PAA).  Reports for other experiments on MBBA
give $\TSC = 45^{\circ}$C and $\TNI = 46^{\circ}$C---see
\cite[Table\,I, p.\,6695]{mkaddem:gartland:00} (note however that the
values for $a_0$, $B$, and $C$ in this table are in error, transcribed
incorrectly from \cite[Table\,1,
p.\,168]{priestly:wojtowicz:sheng:75}).  Thus at room temperature
(taken as $21^{\circ}$C), we would have $\theta = -24$ for
MBBA---while PAA is not even in a liquid crystal phase at room
temperature.

The nematic correlation length has a statistical-physics
interpretation, however in our continuum model, it simply emerges as a
singular-perturbation parameter in the Euler-Lagrange equations
associated with \eqref{eqn:FQ}:
\begin{equation}\label{eqn:EL}
  - L \, \Delta \bfQ +
  \myoverbracket{ \frac{\partial\fb}{\partial\bfQ} } = \bfzero .
\end{equation}
Here $\myoverbracket{~\cdot~}$ denotes the symmetric traceless part:
\begin{equation*}
  \myoverbracket{ \frac{\partial\fb}{\partial\bfQ} } = A \bfQ -
  B \Bigl[ \bfQ^2 - \frac13 \tr\bigl(\bfQ^2\bigr) \bfI \Bigr] +
  C \tr\bigl( \bfQ^2 \bigr) \bfQ .
\end{equation*}
At $T = \TNI$, we have
\begin{equation*}
  - \xiNI^2 \, \Delta \bfQ + \frac1{~\ANI}
  \myoverbracket{ \frac{\partial\fb}{\partial\bfQ} } = \bfzero .
\end{equation*}
When $\bfQ$ is close to a stationary point of $\fb$ (such as $\bfQ =
\bfzero$ or the bottom of a well), then the term $(1/\ANI)
\myoverbracket{ \partial\fb / \partial\bfQ }$ will be close to zero;
otherwise, it will be $O(1)$.  The necessary balance between the terms
in the equation above indicates how $\xiNI$ determines the ``core
size'' of a defect (via a scaling of $X_i = x_i / \xiNI$ for the
``inner solution'').

We have defined $\xiNI$ by linearization around the isotropic state
$\bfQ = \bfzero$, which is always a critical point of $\fb$ and the
solution one would see at the center of an isotropic core.  Note
however that isotropic defect cores are only stable at the high end of
the nematic temperature range (see
\cite{gartland:mkaddem:99,rosso:virga:96}).  The core structure of a
defect is generally more complicated than isotropic disordering, and
this is discussed more in what follows.  In some settings, it is more
natural to define the correlation length by linearizing around the
nontrivial uniaxial $\bfQ$ tensor that gives the global minimum of
$\fb$ for $T < \TNI$ (see for example \cite{ravnik:zumer:09}).
Correlation lengths and core sizes depend on temperature, and this
should be taken into account in non-dimensionalizations if one intends
to explore behavior deep in the nematic phase---we do this below.  In
some situations, a more natural intrinsic length scale is given by a
``biaxial coherence length'' (see
\cite{kralj:virga:zumer:99,penzenstadler:trebin:89})---we consider
this as well in what follows.  The scaling we have adopted above has
been chosen for simplicity.  It is by no means original.  Similar
scalings are employed by all practitioners who work with
Landau-de\,Gennes models.  In addition to references already cited,
see for example \cite{deluca:rey:07,fukuda:stark:yoneya:yokoyama:04,%
  gartland:palffy-muhoray:varga:91,mkaddem:gartland:00,%
  sonnet:kilian:hess:95,ziherl:zumer:97}, among others.

In our rescaled free energy \eqref{eqn:FQbar}, all quantities are
dimensionless; the size of $\Omegabar$ is $O(1)$; and at the high end
of the nematic temperature range, $\Fbar$, $\Qbar$, and $\theta$ are
$O(1)$ as well.  We can see under what circumstances the coupling
coefficient between the elastic and bulk terms is small:
\begin{equation*}
  0 < \xiNIbar \ll 1 ~~ \Leftrightarrow ~~ 0 < \xiNI \ll R ,
\end{equation*}
that is, when the nematic correlation length (isotropic core size) is
small compared to a length scale associated with the problem geometry
(size of $\Omega$).  The dimensionless spatial gradient
$\gradbar\Qbar$ will be $O(1)$ away from defects and $O(1/\xiNIbar)$
in the vicinity of defects.  The limit $\xiNIbar \rightarrow 0$
corresponds to the core size of defects becoming vanishingly small
compared to the size of the problem geometry, with finite-size defects
becoming point or line singularities in the limit.  This can be
thought of as a ``zero-core-size limit'' or a ``large-body limit.''

For the more general functional \eqref{eqn:Feps} (without the
artificially introduced $\eps$), a similar rescaling would give
\begin{subequations}\label{eqn:FQijbar}
\begin{multline}\label{eqn:FQijbara}
  \Fbar[\Qbar] = \int_{\Omegabar} \Bigl[ \frac12 \xiNIbar^2
  \Bigl( \sQbar_{ij,\kbar} \sQbar_{ij,\kbar} +
  \Lbar_2 \sQbar_{ij,\jbar} \sQbar_{ik,\kbar} +
  \Lbar_3 \sQbar_{ij,\kbar} \sQbar_{ik,\jbar} \Bigr) \\
  {} + \frac{\theta}2 \tr \bigl( \Qbar^2 \bigr) -
  \sqrt{3} \tr \bigl( \Qbar^3 \bigr) +
  \frac14 \tr \bigl( \Qbar^2 \bigr){}^{\!2} \Bigr] \text{d} \Vbar ,
\end{multline}
where
\begin{equation}
  \xiNIbar = \frac{~\xiNI}{R} , \quad
  \xiNI := \sqrt{ \! \frac{L_1}{\ANI} } , \quad
  \Lbar_2 := \frac{L_2}{L_1} , \quad
  \Lbar_3 := \frac{L_3}{L_1} .
\end{equation}
\end{subequations}
Here $\Lbar_2$ and $\Lbar_3$ are dimensionless and $O(1)$.  The
definition of the dimensionless parameter $\eps$ in \eqref{eqn:Eepsa}
is not given in \cite{golovaty:montero:14}.  One can assume that it
has been constructed in a way that is analogous to what has been done
here for $\xiNIbar$ in \eqref{eqn:FQbar} and above.

\subsection{Low-temperature limit}

In the ``low temperature'' regime, deep in the nematic phase, several
effects are manifested by the Landau-de\,Gennes models and their
free-energy-minimizing solutions.  These include the degree of
orientational order increasing, the potential wells in $\fb$ becoming
deeper with the barriers between the wells smaller, and correlation
lengths and defect core sizes becoming smaller as well.  These various
features must be taken into account in a good scaling of the
free-energy functional for this range.  The combination of these
features serves to penalize biaxiality less, encouraging localized
biaxiality more as a way for equilibrium tensor fields to avoid the
large free-energy costs of isotropic cores in defects, and this is the
motivation for papers such as
\cite{contreras:lamy:14,majumdar:pisante:henao:15} to analyze this
limit rigorously.

Recall that the only temperature dependence of the parameters in
\eqref{eqn:FQ} is through the $A$ parameter, assumed to have the form
$A = a_0 (T-\TSC)$, $a_0 > 0$, given in \eqref{eqn:AofT}.  Thus $A$
becomes more and more negative, deeper and deeper into the nematic
phase.  The behavior of the scalar order parameter $S$ and bulk
potential $\fb$ in this range can be determined as follows.  Under the
uniaxial assumption \eqref{eqn:uniaxialQ}, one obtains
\begin{equation*}
  \bfQ = S \Bigl( \nhat\otimes\nhat - \frac13 \bfI \Bigr)
  ~ \Rightarrow ~
  \fb(\bfQ) = \frac{A}3 S^2 - \frac{2B}{27} S^3 + \frac{C}9 S^4 =: f(S) ,
\end{equation*}
for which the nontrivial critical points are given by
\begin{subequations}\label{eqn:Spm_fpm_asymp}
\begin{equation}\label{eqn:Spm_asymp}
  \Spm = \frac{B \pm \sqrt{B^2-24AC}}{4C} \approx
  \pm \sqrt{ \! \frac{-3A~}{2C} } , \quad \frac{A}{C} \ll -1 ,
\end{equation}
with associated values
\begin{equation}
  f(\Spm) \approx \frac{-A^2}{4C} , \quad \frac{A}{C} \ll -1 .
\end{equation}
\end{subequations}
The dimensionless ratio $A/C$ is proportional to the reduced
temperature $\theta$ used previously:
\begin{equation*}
  \frac{A}{C} = \frac{\ANI}{C} \frac{A}{\ANI} =
  \frac{\ANI}{C} \theta , \quad
  \theta = \frac{A}{~\ANI} \text{ as in } \eqref{eqn:FQbarb} .
\end{equation*}
Here the ratio $\ANI/C$ is dimensionless and $O(1)$, and we see that
\begin{equation*}
  | \Spm \! | = O\bigl( \! \sqrt{\!-\theta} \bigr) , ~~
  - \frac1{C} f( \Spm \! ) = O\bigl( \theta^2 \bigr) , \quad
  \text{as } \theta \rightarrow -\infty .
\end{equation*}
We will rescale our single-elastic-constant Landau-de\,Gennes model
\eqref{eqn:FQ} for the low-temperature regime using this information
and staying as close as possible to the parameter definitions and
notations used in the preceding subsection.

We scale lengths as before in \eqref{eqn:xbar}, using some length
scale $R$ appropriate to the geometry of the problem domain:
\begin{equation*}
  \xbar_i = \frac{x_i}{R} , \quad
  \Omegabar = \frac1R \Omega .
\end{equation*}
We scale $\bfQ$ and $\calF$ by their low-temperature asymptotic
values, given in \eqref{eqn:Spm_fpm_asymp}:
\begin{equation}\label{eqn:QtildeFtilde}
  \bfQ =  \sqrt{ \! \frac{-A~}{C} } \, \Qtilde , \quad
  \calF = \frac{~~A^2}{C} R^3 \Ftilde .
\end{equation}
Introducing these scalings into \eqref{eqn:FQ} and simplifying, we
obtain
\begin{subequations}\label{eqn:FQtilde}
\begin{equation}\label{eqn:FQtildea}
  \Ftilde[\Qtilde] = \int_{\Omegabar} \Bigl[
  \frac12 \eps^2 \xiNIbar^2 \bigl|\gradbar\Qtilde\bigr|{}^2 -
  \frac12 \tr\bigl(\Qtilde^2\bigr) -
  \sqrt{3} \, \eps \tr\bigl(\Qtilde^3\bigr) +
  \frac14 \tr\bigl(\Qtilde^2\bigr){}^{\!2} \Bigr] \text{d}\Vbar ,
\end{equation}
where
\begin{equation}
  \eps := \frac1{\sqrt{\!-\theta}} , \quad
  \theta = \frac{A}{~\ANI} = \frac{T-\TSC}{\TNI-\TSC} , \quad
  \xiNIbar = \frac{~\xiNI}{R} , \quad
  \xiNI = \sqrt{ \! \frac{L}{~\ANI} } .
\end{equation}
\end{subequations}
Here the definitions of $\theta$, $\xiNI$, and $\xiNIbar$ are exactly
as before, and we have adopted the small parameter $\eps$ as a
control parameter:
\begin{equation*}
  \theta \ll - 1
  ~~ \leftrightarrow ~~
  0 < \eps \ll 1 .
\end{equation*}
We have deliberately written the temperature-dependent nematic
correlation length, $\xin$, in terms of $\eps$ and the nematic
correlation length at $T=\TNI$ ($\xiNI$, defined in
\eqref{eqn:FQbarb}):
\begin{equation}\label{eqn:xin}
  \xin := \sqrt{\!\frac{L}{-A~}} =
  \sqrt{\!\frac{~\ANI}{-A~}} \sqrt{\!\frac{L}{~\ANI}} =
  \frac1{\sqrt{\!-\theta}} \, \xiNI =
  \eps \, \xiNI .
\end{equation}
The relevance of $\xin$ can already be gleaned from the coupling
between the terms $\xiNIbar^2 | \gradbar\Qbar |^2$ and $\theta \tr
(\Qbar)^2$ in \eqref{eqn:FQbara}.  We can see above how the
correlation length scales with temperature, shrinking like
$1/\sqrt{\!-\theta}$ as $\theta$ becomes more and more negative.  In
\eqref{eqn:FQtildea}, we have again used the relation $\ANI = B^2 \! /
27C$ from \eqref{eqn:temps} to simplify the coefficient of the
$\tr(\bfQ^3)$ term.

The relationship between the ``large body'' scaling, in
\eqref{eqn:FQbar} and \eqref{eqn:FQijbar}, and the ``low temperature''
scaling in \eqref{eqn:FQtilde} is given by
\begin{equation}\label{eqn:barvstilde}
  \Qbar = \sqrt{\!-\theta} \, \Qtilde , \quad
  \Fbar = \theta^2 \Ftilde ,
\end{equation}
which can be verified directly from the definitions of $\Qbar$,
$\Fbar$, $\Qtilde$, $\Ftilde$, and $\theta$, with the help of $\ANI =
B^2\!/27C$ from \eqref{eqn:temps}.  One can, in fact, derive
\eqref{eqn:FQtilde} directly from \eqref{eqn:FQbar} by introducing the
relations \eqref{eqn:barvstilde} into \eqref{eqn:FQbar} and
simplifying.  We note that \eqref{eqn:FQbar} and \eqref{eqn:FQijbar}
are not good scalings for exploring the low-temperature limit ($\theta
\rightarrow -\infty$), because they do not take into account the
behavior of $|\Qbar|$ in that range:
\begin{equation*}
  \tr \bigl( \Qbar^2 \bigr) = O(|\theta|) , ~~
  \tr \bigl( \Qbar^3 \bigr) = O\bigl(|\theta|^{3/2}\bigr) , ~~
  \tr \bigl( \Qbar^2 \bigr){}^{\!2} = O\bigl(|\theta|^2\bigr) , ~~
  \text{as } \theta \rightarrow - \infty .
\end{equation*}

In our free-energy functional rescaled for the low-temperature regime,
\eqref{eqn:FQtilde}, all quantities are again dimensionless; the size
of $\Omegabar$ is again $O(1)$; and for $\theta \le -1$ ($0 < \eps \le
1$), say, $\Ftilde$ and $\Qtilde$ are $O(1)$ as well.  The
non-dimensionalized model shares many features with the functional
\eqref{eqn:FQbar}, which is scaled for the large-body regime.  Both
are left with the same two dimensionless parameters, $\xiNIbar$ (the
ratio of the nematic correlation length at $T=\TNI$ to the geometric
length scale $R$) and the reduced temperature $\theta$ (which enters
\eqref{eqn:FQtilde} through $\eps = 1 / \sqrt{\!-\theta}$\,).  In
\eqref{eqn:FQbar}, $\theta$ can be positive, negative, or zero, which
allows for using that scaling up to and including the co-existence
temperature range ($0 \le \theta \le 9/8$); while in
\eqref{eqn:FQtilde}, $\theta < 0$ is assumed---the two scaled models
are identical, in fact, when $\theta = -1$ (for any $\xiNIbar$).  The
models share the feature of a dimensionless coupling coefficient
between the elastic and the bulk terms ($\xiNIbar^2$ in
\eqref{eqn:FQbar}, $\eps^2\xiNIbar^2$ in \eqref{eqn:FQtilde}), and
these coupling coefficients are expected to be quite small in the
respective regimes of interest.  Both the large-body limit explored in
\cite{bauman:park:phillips:12,golovaty:montero:14,majumdar:zarnescu:10}
($\xiNIbar \rightarrow 0$) and the low-temperature limit in
\cite{contreras:lamy:14,majumdar:pisante:henao:15} ($\eps \rightarrow
0$) correspond to circumstances in which the (temperature dependent)
core size of defects becomes vanishingly small compared to the size of
the domain---a ``zero-relative-core-size limit.''

The presence of the factor $\eps$ in the coefficient of the $\tr
(\Qtilde^3)$ term in \eqref{eqn:FQtilde},
\begin{equation}\label{eqn:fbtilde}
  \fbepstilde(\Qtilde) = - \frac12 \tr\bigl(\Qtilde^2\bigr) -
  \sqrt{3} \, \eps \tr\bigl(\Qtilde^3\bigr) +
  \frac14 \tr\bigl(\Qtilde^2\bigr){}^{\!2} ,
\end{equation}
drives that term to zero in the low-temperature limit and leaves the
rescaled bulk ordering potential in the form
\begin{equation*}
  \fbztilde(\Qtilde) = - \frac12 \tr\bigl(\Qtilde^2\bigr) +
  \frac14 \tr\bigl(\Qtilde^2\bigr){}^{\!2} .
\end{equation*}
Thus $\fbztilde$ is expressed only in terms of even powers,
$\tr\bigl(\Qtilde^2\bigr)$ and $\tr\bigl(\Qtilde^2\bigr){}^{\!2}$, and
more closely resembles the types of Landau expansions found in the
Ginzburg-Landau theory of superconductivity and the Landau-Lifshitz
theory of ferromagnetism.  The potential $\fbztilde$ has a local
maximum of $\fbztilde=0$ at the isotropic state $\Qtilde=\bfzero$ and
absolute minima given by
\begin{equation*}
  \min \fbztilde = - \frac14 , ~~ \text{when }
  \tr\bigl(\Qtilde^2\bigr) = 1 .
\end{equation*}
The distinguishing feature here is that the set of bulk-minimizing
tensor order parameters ($\Qtilde$ satisfying $\tr(\Qtilde^2)=1$)
includes a continuum of biaxial states (in addition to uniaxial states
of the form \eqref{eqn:uniaxialQ}).  Thus the usual penalty from $\fb$
for biaxiality vanishes in this limit.  This behavior has been known
for some time (see \cite{lyuksyutov:78}).  It is the basis of the
so-called ``Lyuksyutov constraint'' ($\tr(\bfQ^2) = \text{const}$),
which has been used on occasion to obtain approximations from
Landau-de\,Gennes models deep in the nematic phase---see for example
\cite{kralj:virga:zumer:99,lyuksyutov:78,penzenstadler:trebin:89}.
The justification of the Lyuksyutov constraint is well illustrated by
\eqref{eqn:fbtilde}, and this behavior also underlies the analyses in
\cite{contreras:lamy:14,majumdar:pisante:henao:15}.

The main thrust of \cite{contreras:lamy:14} is to prove that at
sufficiently low temperatures, tensor fields that minimize the
Landau-de\,Gennes free energy \eqref{eqn:FQa} do not have isotropic
cores (even if boundary conditions dictate that a defect of some kind
must be present).  Such tensor fields avoid the large free-energy cost
of such cores by going through localized biaxial transitions instead.
The paper \cite{majumdar:pisante:henao:15} provides some
generalizations of \cite{contreras:lamy:14} and focuses some
consideration on the ``vanishing-elasticity limit'' aspect.  Both
papers use some rescaling but leave the free-energy functionals in
partly dimensional form: the main integrals ((10) in
\cite{contreras:lamy:14}, (14) in \cite{majumdar:pisante:henao:15})
have the physical dimensions of a volume and the coupling coefficients
are lengths${}^2$\!.  One can establish some connections with the
low-temperature scaling we have adopted here in \eqref{eqn:FQtilde}.
Both \cite{contreras:lamy:14} and \cite{majumdar:pisante:henao:15}
scale $\bfQ$ by the temperature-dependent scalar order parameter
$\sqrt{2/3}\,S_{+}$, whereas here we have scaled $\bfQ$ by the
asymptotic value of this,
\begin{equation*}
  \sqrt{\frac23} \, S_{+} \approx \sqrt{\!\frac{-A~}{C}} ,
  ~~ \text{for } \frac{A}{C} \ll -1 ,
\end{equation*}
from \eqref{eqn:Spm_asymp} and \eqref{eqn:QtildeFtilde}.  The reduced
temperatures, denoted by $t$ in both papers (but defined slightly
differently in each), are related to $\theta$ in \eqref{eqn:FQbarb}
here via
\begin{equation*}
  t = - \frac1{27} \theta ~~
  \text{in \cite{contreras:lamy:14},} \quad
  t = - \theta ~~
  \text{in \cite{majumdar:pisante:henao:15},}
\end{equation*}
and the coupling coefficients are related to $\xiNI$ in
\eqref{eqn:FQbarb} here by
\begin{equation*}
  \widetilde{L} = \frac19 \xiNI^2 ~~
  \text{in \cite{contreras:lamy:14},} \quad
  \Lbar = \frac12 \xiNI^2 ~~
  \text{in \cite{majumdar:pisante:henao:15}.}
\end{equation*}
In both papers, the rescaled integral functionals are parametrized by
the large parameter $t$, rather than the small parameter $\eps$ we
have used here.  If one were to scale the integral (10) in
\cite{contreras:lamy:14} or (14) in \cite{majumdar:pisante:henao:15}
by $1/t$ and non-dimensionalize the lengths, then one would arrive at
something close to \eqref{eqn:FQtilde} here.  The scaling that we have
adopted here also enables us to make contact with some aspects of
biaxiality in the low-temperature regime, which we discuss next.

\subsection{Biaxiality in the low-temperature regime}

The biaxial nature of defect cores in typical temperature ranges (as
modeled in the Landau-de\,Gennes framework) was demonstrated in
\cite{schopohl:sluckin:87}, based upon earlier predictions.  The size
of such ``biaxial cores'' is known to be determined by the
(temperature-dependent) ``biaxial coherence length,'' defined in
\cite{kralj:virga:zumer:99,penzenstadler:trebin:89} as
\begin{equation*}
  \xib := \sqrt{\!\frac{L}{~B\Splus}} .
\end{equation*}
In the low-temperature regime, this can be related to parameters we
have used elsewhere in this note:
\begin{equation*}
  \Splus \approx \sqrt{\!\frac{-3A~}{2C}} , ~~ \frac{A}{C} \ll -1
  ~~ \Rightarrow ~~
  \xib = \sqrt{\!\frac{L}{~B\Splus}} \approx
  \frac13 2^{1/4} \eps^{1/2} \xiNI , ~~ 0 < \eps \ll 1 .
\end{equation*}
Observe the difference in the temperature dependence of $\xib$ versus
that of the nematic correlation length $\xin$:
\begin{equation*}
  \xib = \frac13 2^{1/4} \eps^{1/2} \xiNI \bigl( 1 + O(\eps) \bigr) ,
  ~~ \text{as } \eps \rightarrow 0+
  ~~~ \text{vs} ~~~
  \xin = \eps \, \xiNI , ~~ \text{from } \eqref{eqn:xin} .
\end{equation*}
At high temperatures, however, these lengths are essentially
equivalent:
\begin{equation*}
  \ANI = \frac{~B^2}{27C} , ~ \SNI = \frac{B}{3C}
  ~~ \Rightarrow ~~
  \xin = \sqrt{\!\frac{L}{~\ANI}} =
  3 \sqrt{\!\frac{L}{~B\SNI}} = 3 \xib , ~
  \text{at } T = \TNI .
\end{equation*}
We will see below that the biaxial coherence length can be identified
from a balance in the equilibrium Euler-Lagrange equations for $\bfQ$
for the ``inner solution'' in a biaxial defect core.

There is a subtle mechanism by which biaxiality can enable defects
that cost less free energy than isotropic cores.  This has been
explored in several papers, including
\cite{deluca:rey:07,gartland:mkaddem:99,kralj:virga:zumer:99,%
  mkaddem:gartland:00,palffy:gartland:kelly:94,penzenstadler:trebin:89,%
  schopohl:sluckin:87,sonnet:kilian:hess:95}.  It can be loosely
described as follows.  If one fixes the eigenframe of the $\bfQ$
tensor ($\ehat_1$, $\ehat_2$, $\ehat_3$, say) and studies the bulk
free-energy surface of $\fb$ as a function of the two independent
eigenvalues of $\bfQ$, then one finds three wells, at positively
ordered uniaxial states of the form \eqref{eqn:uniaxialQ} with $S =
\Splus > 0$ ($\Splus$ as given in \eqref{eqn:Spm_asymp}), each well
corresponding to a different director $\nhat = \ehat_1$, $\ehat_2$,
$\ehat_3$.  These three global minima are separated from each other by
saddle points, which also correspond to uniaxial states but which are
negatively ordered, with $S = \Sminus < 0$ (again from
\eqref{eqn:Spm_asymp}).  An excursion by the eigenvalues of $\bfQ$
from one well to another thus enables a reorientation of $\nhat$
without rotating the eigenframe of $\bfQ$.  This mechanism is
sometimes referred to as ``eigenvalue exchange,'' and this is the
mechanism at work in the core of a strength-1/2 disclination line and
a hybrid cell at extremely narrow cell gap (see
\cite{kralj:virga:zumer:99,palffy:gartland:kelly:94,%
  schopohl:sluckin:87,sonnet:kilian:hess:95}).  In geometry that
encourages a point defect, the defect core will instead form a small
ring or torus across which such a biaxial transition occurs
\cite{deluca:rey:07,gartland:mkaddem:99,kralj:virga:zumer:99,%
  mkaddem:gartland:00,penzenstadler:trebin:89,sonnet:kilian:hess:95}.
With the exception of the wells and the saddle points, all other
points along the path of such an excursion would be biaxial (and very
strongly so along certain segments).

A significant component of the cost of such a ``biaxial escape'' is
climbing the potential barrier to cross the ``mountain pass'' between
two wells.  In the low-temperature regime, however, the magnitude of
this potential barrier becomes smaller, which can be seen as follows.
In the low-temperature scaling \eqref{eqn:fbtilde} (with $\Qtilde$ as
defined in \eqref{eqn:QtildeFtilde}), the critical scalar order
parameters (rescaled analogues of \eqref{eqn:Spm_asymp} for
$\fbepstilde (\Qtilde)$) are given by
\begin{equation*}
  \Stildepm = \frac{3\sqrt{3}\,\eps \pm \sqrt{27\eps^2+24}}4 =
  \pm \sqrt{\frac32} + \frac{3\sqrt{3}}4\,\eps + O\bigl(\eps^2\bigr) .
\end{equation*}
Denoting
\begin{equation*}
  \Qtildepm = \Stildepm \Bigl( \nhat\otimes\nhat - \frac13 \bfI \Bigr) ,
\end{equation*}
we obtain
\begin{equation*}
  \fbepstilde\bigl(\Qtildepm\!\bigr) =
  - \frac14 \mp \frac1{\sqrt{2}~} \eps + O\bigl(\eps^2\bigr)
  \quad \text{and} \quad
  \tr \bigl( \Qtildepm^2 \! \bigr) =
  1 \pm \frac3{\sqrt{2}~} \eps + O\bigl(\eps^2\bigr) .
\end{equation*}
With the eigenvalues of $\Qtilde$ taking the path of least resistance
along such an excursion between wells, we would then have
\begin{subequations}
\begin{equation}\label{eqn:fbepstildepm}
  \fbepstilde\bigl(\Qtildeplus\!\bigr) \le
  \fbepstilde\bigl(\Qtilde\bigr) \le
  \fbepstilde\bigl(\Qtildeminus\!\bigr) , \quad
  \fbepstilde\bigl(\Qtildeminus\!\bigr) -
  \fbepstilde\bigl(\Qtildeplus\!\bigr) =
  \sqrt{2} \, \eps + O\bigl(\eps^2\bigr)
\end{equation}
and
\begin{equation}\label{eqn:trQtilde}
  \tr\bigl(\Qtildeminus^2\!\bigr) \le
  \tr\bigl(\Qtilde^2\bigr) \le
  \tr\bigl(\Qtildeplus^2\!\bigr)
  ~~ \Rightarrow ~~
  \bigl| \tr\bigl(\Qtilde^2\bigr) - 1 \bigr| \le
  \frac3{\sqrt{2}~} \, \eps + O\bigl(\eps^2\bigr) .
\end{equation}
\end{subequations}
The $O(\eps)$ potential barrier seen in \eqref{eqn:fbepstildepm} is
much more favorable than the cost of isotropic melting:
$\fbepstilde(\bfzero) - \fbepstilde(\Qtildeplus) = 1/4 + O(\eps)$.  A
smooth $O(\eps)$ change in $\fbepstilde$ (consistent with
\eqref{eqn:fbepstildepm}) over the course of an $O(1)$ change in
$\Qtilde$ implies $\partial\fbepstilde / \partial\Qtilde = O(\eps)$
along such a path, and this can be seen from the Euler-Lagrange
equations in this scaling:
\begin{equation*}
  - \eps^2 \xiNIbar^2 \, \overline{\Delta} \Qtilde +
  \bigl[ \tr\bigl(\Qtilde^2\bigr) - 1 \bigr] \Qtilde -
  3 \sqrt{3} \, \eps \Bigl[ \Qtilde^2 -
  \frac12 \tr\bigl(\Qtilde^2\bigr) \bfI \Bigr] = \bfzero .
\end{equation*}
The middle and last terms above are seen to be $O(\eps)$ (with the
help of \eqref{eqn:trQtilde}), and the necessary balance that must
exist in the ``interior layer'' between these terms and the leading
term above gives a length scale in the biaxial core of $\eps^{1/2}
\xiNIbar$ (which is the low-temperature asymptotic value of
$\xibbar$).  This dependence on reduced temperature ($\eps^{1/2} =
(-\theta)^{-1/4}$) is consistent with results in
\cite{contreras:lamy:14,majumdar:pisante:henao:15}, where some of
these same quantities of interest are considered in a more rigorous
way.  While biaxial cores are somewhat larger than isotropic cores and
shrink more slowly than isotropic cores as temperature is reduced,
they still become vanishingly small in either the large-body limit or
the low-temperature limit, as do the biaxial rings and tori.

\section{Conclusions}

The Landau-de\,Gennes model is a mesoscopic model that contains
intrinsic length scales of molecular order associated with features
such as core sizes of point defects and disclination lines.  The
Oseen-Frank model, on the other hand, is a macroscopic model and
contains no such intrinsic length scales: the defects of equilibrium
director fields in Oseen-Frank models are point or line singularities.
The scaling analysis in Sec.\,\ref{sec:analysis} shows that the limits
explored in \cite{bauman:park:phillips:12,contreras:lamy:14,%
  golovaty:montero:14,majumdar:pisante:henao:15,majumdar:zarnescu:10}
concern the passage from a mesoscopic model to a macroscopic model as
the geometric length scales become large compared to the intrinsic
length scales.  In such limits, core sizes (isotropic or biaxial)
become zero.  The situation is analogous to a ``London limit'' for
Ginzburg-Landau models, in which the role of the Ginzburg-Landau
parameter is here played by the ratio of an intrinsic length scale to
a geometric one---a proper scaling and non-dimensionalization of the
model are required to identify this.  Such limits are also analogous
to the ``large body limit'' in the Landau-Lifshitz theory of
ferromagnetism \cite{desimone:93} and (properly interpreted) can
indeed be thought of as ``the Oseen-Frank limit,'' as in the title of
\cite{majumdar:zarnescu:10}.

The analysis in the papers \cite{bauman:park:phillips:12,%
  contreras:lamy:14,majumdar:pisante:henao:15,majumdar:zarnescu:10}
remains valid after a re-interpretation is done of the problem
parameters (into appropriate dimensionless forms)---the functionals in
\eqref{eqn:FQa} versus \eqref{eqn:FQbara} and \eqref{eqn:Fepsa} versus
\eqref{eqn:FQijbara} involve the same terms in a formal sense, just
differently scaled.  One should take heed, however, of the pitfalls of
analyzing such a physical model in fully (or partially) dimensional
form.  The numerical values of quantities of different physical
dimensions can change their relative sizes when the system of units is
changed.  It is also the case that familiar Sobolev-type norms, such
as
\begin{equation*}
  \| \bfQ \|_1^2 = \int_{\Omega} \bigl[
  | \nabla \bfQ |^2 + | \bfQ |^2 \bigr] \, \text{d}V
\end{equation*}
(which are used in most of the papers in this area), can't even be
used in this form unless lengths have been non-dimensionalized;
otherwise the first term in the integrand above would have dimensions
of $1/\text{length}^2$\!, while the second term would be dimensionless
(and the two terms couldn't be combined).  We acknowledge that
\cite{davis:gartland:98} is guilty of this error as well.  A remedy
for this would be to use weighted Sobolev norms, such as
\begin{equation*}
  \| \bfQ \|_1^2 = \int_{\Omega} \bigl[
  w_1 | \nabla \bfQ |^2 + w_2 | \bfQ |^2 \bigr] \, \text{d}V ,
\end{equation*}
with $w_1$ and $w_2$ chosen with dimensions to render the combination
meaningful---the choice of weights in such a norm would be problem
dependent, however, which is not desirable.

An additional difficulty of attempting to analyze such limits in
dimensional form is that as $L \rightarrow 0$ in \eqref{eqn:FQa}, one
quickly gets beyond the measured values of $L$ for real liquid crystal
materials.  Properly non-dimensionalized models such as
\eqref{eqn:FQbar}, \eqref{eqn:FQijbar}, and \eqref{eqn:FQtilde},
however, do not suffer from this difficulty: the smallness of the
coupling coefficients $\xiNIbar^2$ in \eqref{eqn:FQbara} and
\eqref{eqn:FQijbara} and $\eps^2\xiNIbar^2$ in \eqref{eqn:FQtildea} is
related only to the reduced temperature and to the ratio of the core
size to the size of the problem domain (and remains within the bounds
of reasonable physical values over a broad range).

While the analyses in \cite{bauman:park:phillips:12,%
  contreras:lamy:14,golovaty:montero:14,majumdar:pisante:henao:15,%
  majumdar:zarnescu:10} are modeled after earlier work by others on
models with some similar features (such as Ginzburg-Landau), technical
challenges accompany the analysis of Landau-de\,Gennes models by
virtue of the tensorial nature of the state variable and the multiple
terms and parameters and the complexity of the functional.  An effort
to address some of the issues we have taken up here was made in a
brief appendix in \cite{nguyen:zarnescu:13}.  We note that both of
these limits (``large body'' and ``low temperature'') are
idealizations from a physical point of view.  When the size of the
domain containing the liquid crystal becomes too large (compared to
intrinsic length scales), thermal fluctuations would wash out any
orientational order that other factors might try to induce.  Also, at
sufficiently low temperatures, all liquid crystal materials would
eventually crystallize or transition to a glass or to a non-nematic
liquid crystal phase (and both the Landau-de\,Gennes and Oseen-Frank
models would no longer be valid).  The study of such limits, however,
can give useful insights.

One of the interesting aspects of the analyses done in the papers
\cite{bauman:park:phillips:12,contreras:lamy:14,%
  golovaty:montero:14,majumdar:pisante:henao:15,majumdar:zarnescu:10}
is that they provide, to some extent, a justification for using the
Oseen-Frank model to compute equilibrium director fields in the case
when line disclinations are present (in the large-body or
low-temperature regime): even though the Oseen-Frank free energy of
such solutions is infinite, the equilibrium director fields (found
from the Euler-Lagrange equations) are the limiting director fields
associated with solutions of a Landau-de\,Gennes model, which
solutions have finite free energy for all positive values of the
coupling coefficients, though diverging in the limit as the coupling
coefficients (as we have written them) approach zero.  However, the
divergence of the Oseen-Frank free-energy functional $F$ evaluated on
any director field containing a disclination line prevents it from
being used to assess local or global stability of such solutions.

In all but the simplest of settings, one must resort to numerical
methods to approximate solutions of Landau-de\,Gennes or Oseen-Frank
models, and this sometimes introduces aspects that are closely related
to issues we have highlighted here.  An example of this is the use of
``penalty methods'' for Oseen-Frank models (or for the related
Ericksen-Leslie hydrodynamic equations)---see
\cite{adler:emerson:maclachlan:manteuffel:16,walkington:11}.  This
popular approach avoids imposing the constraint $|\nhat|=1$ pointwise
(and the necessary introduction of an associated Lagrange-multiplier
field) by instead adding a term to the free-energy functional that
penalizes departures from $|\nhat|=1$.  If the model \eqref{eqn:OZF}
were to be left in dimensional form, then the penalized version of it
could be written
\begin{equation*}
  \Fxi[\nhat] = \frac{K}2 \int_{\Omega} \Bigl[ | \nabla\nhat |^2 +
  \xi^{-2} \bigl( |\nhat|^2 - 1 \bigr){}^{\!2} \Bigr] \text{d}V .
\end{equation*}
The purely numerical ``penalty parameter'' $\xi$ must have the
physical dimensions of a length, and the macroscopic model now has the
complexion of a mesoscopic model.  The smaller $\xi$ is (compared to
the size of the domain, say), the greater the extent to which the
constraint is imposed.  Equilibria of $\Fxi$ that possess defects
would have finite ``cores'' of size $O(\xi)$ (in which the length of
$\nhat$ would depart significantly from one).  The non-dimensionalized
version of the penalized functional above (lengths scaled by $R$,
energy by $KR$) would take the form
\begin{equation*}
  \Fbarxibar[\nhat] =
  \frac12 \int_{\Omegabar} \Bigl[ \bigl| \gradbar\nhat \bigr|{}^2 +
  \xibar^{-2} \bigl( |\nhat|^2 - 1 \bigr){}^{\!2} \Bigr] \text{d}\Vbar ,
\end{equation*}
which in two space dimensions would be precisely the Ginzburg-Landau
functional as studied in \cite{bethuel:brezis:helein:93,%
  bethuel:brezis:helein:94}.  All of the Ginzburg-Landau analytical
machinery would apply and give analogous strong convergence results
(away from singular sets) for minimizers of $\Fbarxibar$, as the
dimensionless penalty parameter $\xibar \rightarrow 0$.  Less appears
to be known in the setting of dynamics (liquid crystal director
dynamics or full Ericksen-Leslie hydrodynamics).  Discretization of
such problems (via finite elements or finite differences, say) would
introduce another length scale into the discretized model, the mesh
size or grid size (typically denoted by $h$), and one would need to
take into consideration different regimes of dimensionless ratios of
$h$, $\xi$, and $R$.

Landau-de\,Gennes and Oseen-Frank models contain a large number of
physical parameters, and parameter-dependent effects (changes of state
of solutions, bifurcations, structural phase transitions, behaviors in
extreme ranges of parameters) are of significant importance in
general.  Dimensional analysis, scaling, and balances (as used in
boundary/interior-layer theory) provide a useful means to understand
and interpret some of these behaviors.  They have enabled us here to
determine the proper interpretation of ``vanishing-elasticity limits''
and to see one way that length scales for isotropic cores and biaxial
cores can be deduced.  These simple tools provide a useful complement
to approaches coming from the realm of mathematical analysis.

\begin{acknowledgements}
  We are grateful to several individuals for feedback on a first
  version of this note \cite{gartland:15}, in particular to John Ball,
  Apala Majumdar, Andr\'{e} Sonnet, Noel Walkington, and Arghir
  Zarnescu.
\end{acknowledgements}

\end{document}